\documentclass[final,5p,times,twocolumn]{elsarticle}

\usepackage{amsmath}
\usepackage{amssymb}
\usepackage{graphicx}
\usepackage{color}
\usepackage{hyperref}

\usepackage[mathscr,scaled=1.15]{urwchancal}
\DeclareFontFamily{OT1}{pzc}{}
\DeclareFontShape{OT1}{pzc}{m}{it}%
{<-> s * [1.15] pzcmi7t}{}
\DeclareMathAlphabet{\mathpzc}{OT1}{pzc}{m}{it}

\definecolor{purple}{rgb}{0.5,0,0.5}
\definecolor{blue}{rgb}{0.0,0,0.9}

\biboptions{sort&compress}

\journal{Physics Letters B}

\begin{document}

\begin{frontmatter}

\title{Flavour symmetry breaking in the kaon parton distribution amplitude}

\author[NanjingMOE,Nanjing]{Chao Shi}
\author[UA]{Lei Chang}
\author[ANL]{Craig D. Roberts}
\author[JARA]{Sebastian M. Schmidt}
\author[KSU]{Peter C. Tandy}
\author[Nanjing,Nanjing2,Nanjing3]{Hong-shi Zong}

\address[NanjingMOE]{Key Laboratory of Modern Acoustics, MOE, Institute of Acoustics, Nanjing University, Nanjing 210093, China}

\address[Nanjing]{Department of Physics, Nanjing University, Nanjing 210093, China}
\address[UA]{CSSM, School of Chemistry and Physics
University of Adelaide, Adelaide SA 5005, Australia}
\address[ANL]{Physics Division, Argonne National Laboratory, Argonne, Illinois 60439, USA}
\address[KSU]{Center for Nuclear Research, Department of
Physics, Kent State University, Kent, Ohio 44242, USA}
\address[JARA]{Institute for Advanced Simulation, Forschungszentrum J\"ulich and JARA, D-52425 J\"ulich, Germany}
\address[Nanjing2]{State Key Laboratory of Theoretical Physics, Institute of Theoretical Physics, CAS, Beijing 100190, China}
\address[Nanjing3]{Joint Center for Particle, Nuclear Physics and Cosmology, Nanjing 210093, China}

\date{13 May 2014}

\begin{abstract}
$\,$\\[-7ex]\hspace*{\fill}{\emph{Preprint no}. ADP-13-21/T876}\\[1ex]
We compute the kaon's valence-quark (twist-two parton) distribution amplitude (PDA) by projecting its Poincar\'e-covariant Bethe-Salpeter wave-function onto the light-front.  At a scale $\zeta=2\,$GeV, the PDA is a broad, concave and asymmetric function, whose peak is shifted 12-16\% away from its position in QCD's conformal limit.  These features are a clear expression of SU$(3)$-flavour-symmetry breaking.  They show that the heavier quark in the kaon carries more of the bound-state's momentum than the lighter quark and also that emergent phenomena in QCD modulate the magnitude of flavour-symmetry breaking: it is markedly smaller than one might expect based on the difference between light-quark current masses.  Our results add to a body of evidence which indicates that at any energy scale accessible with existing or foreseeable facilities, a reliable guide to the interpretation of experiment requires the use of such nonperturbatively broadened PDAs in leading-order, leading-twist formulae for hard exclusive processes instead of the asymptotic PDA associated with QCD's conformal limit.  We illustrate this via the ratio of kaon and pion electromagnetic form factors: using our nonperturbative PDAs in the appropriate formulae, $F_K/F_\pi=1.23$ at spacelike-$Q^2=17\,$GeV$^2$, which compares satisfactorily with the value of $0.92(5)$ inferred in $e^+ e^-$ annihilation at $s=17\,$GeV$^2$.
\end{abstract}

\begin{keyword}
dynamical chiral symmetry breaking \sep
Dyson-Schwinger equations \sep
factorisation in heavy-meson decays \sep
flavour symmetry breaking \sep
light pseudoscalar mesons \sep
parton distribution amplitudes \sep
strange quarks

\smallskip

\emph{Preprint no}. ADP-13-21/T841



\end{keyword}

\end{frontmatter}

\noindent\textbf{1.$\;$Introduction}.
Kaons are strong-interaction bound-states defined by their valence-quark content: a $\bar u$- or $\bar d$-quark combined with the $s$-quark, or the opposite antiparticle-particle combination.  The current-mass of the $u/d$-valence-quark is truly light but that of the $s$-quark has a value commensurate with $\Lambda_{\rm QCD}$, QCD's dynamically-generated mass-scale. 
As we shall describe, this marked imbalance between current-masses provides at least two compelling reasons for studying kaons.  However, given that the $s$-quark is neither light nor heavy, elucidating the impact of the imbalance is challenging because it requires the use of nonperturbative techniques within QCD.

The first thing one would like to explore originates in the observation that with the introduction of the quark model as a classification scheme for the hadron spectrum \cite{GellMann:1964nj,Zweig:1981pd} it became common to assume, in the absence of reliable dynamical information to the contrary, that hadron wave functions and interaction currents exhibit SU$(2)\,\otimes\,$SU$(3)$ spin-flavour symmetry.  That assumption has implications for numerous observables, including the hadron spectrum itself and a host of other static and dynamical properties.  Moreover, in an asymptotically free gauge field theory with $N_c$ colours, this symmetry is exact on $1/N_c \simeq 0$ \cite{Jenkins:2001it}.  Kaons therefore provide the simplest system in which the accuracy of these assumptions and predictions can be tested.

The second aspect convolves the first challenge with the fact that, as strong interaction bound states whose decay is mediated only by the weak interaction, so that they have a relatively long lifetime, kaons have been instrumental in establishing the foundation and properties of the Standard Model; notably, the physics of CP violation.  In this connection the nonleptonic decays of $B$ mesons are crucial because, e.g., the transitions $B^{\pm}\to (\pi K)^{\pm}$ and $B^{\pm} \to \pi^{\pm} \pi^0$ provide access to the imaginary part of the CKM matrix element $V_{ub}$: $\gamma = {\rm Arg}(V_{ub}^\ast)$ \cite{Neubert:1998jq}.  Factorisation theorems have been derived and are applicable to such decays  \cite{Beneke:2001ev}.  However, the formulae involve a certain class of so-called ``non-factorisable'' corrections because the parton distribution amplitudes (PDAs) of strange mesons are not symmetric with respect to quark and antiquark momenta.  Therefore, any derived estimate of $\gamma$ is only as accurate as the evaluation of both the difference between $K$ and $\pi$ PDAs and also their respective differences from the asymptotic distribution, $\varphi^{\rm asy}(u)=6 u (1-u)$.  Amplitudes of twist-two and -three are involved.  With this motivation, we focus on the twist-two amplitudes herein.

Historically, the difficulty with placing constraints on this sort of nonfactorisable contribution is that methods such as lattice gauge theory, QCD sum rules and large-$N_c$ provide little information about the QCD dynamics relevant to hadronic $B$-decays.  We therefore employ QCD's Dyson-Schwinger equations, whose value in the computation of valence-quark distribution amplitudes has recently been established \cite{Chang:2013pqS,Cloet:2013tta,Chang:2013epa,Chang:2013nia,Segovia:2013ecaS}.

One of the key features to emerge from Refs.\,\cite{Chang:2013pqS,Cloet:2013tta,Chang:2013epa,Chang:2013nia,Segovia:2013ecaS} is the crucial role played by dynamical chiral symmetry breaking (DCSB) in shaping PDAs.  DCSB is a remarkable emergent feature of the Standard Model.  It plays a critical role in forming the bulk of the visible matter in the Universe \cite{national2012Nuclear} and is expressed in numerous aspects of the spectrum and interactions of hadrons; e.g., the large splitting between parity partners \cite{Chang:2009zb,Chang:2011ei,Chen:2012qr} and the existence and location of a zero in some hadron elastic and transition form factors \cite{Wilson:2011aa,Cloet:2013gva}.  The impact of DCSB is expressed with particular force in properties of light pseudoscalar mesons.  Indeed, their very existence as the lightest hadrons is grounded in DCSB.

\medskip

\noindent\textbf{2.$\;$Computing the kaon twist-two PDA}.
The kaon's valence-quark distribution amplitude may be obtained via
\begin{equation}
f_K\, \varphi_K(u) = N_c {\rm tr}\,
Z_2 \! \int_{dq}^\Lambda \!\!
\delta(n\cdot q_\eta - u \,n\cdot P) \,\gamma_5\gamma\cdot n\, \chi_K^P(q_\eta,q_{\bar\eta})\,,
\label{kaonPDA}
\end{equation}
where: $N_c=3$; $f_K$ is the kaon's leptonic decay constant; the trace is over spinor indices; $\int_{dq}^\Lambda$ is a Poincar\'e-invariant regularization of the four-dimensional integral, with $\Lambda$ the ultraviolet regularization mass-scale; $Z_{2}(\zeta,\Lambda)$, with $\zeta$ the renormalisation scale, is the quark wave-function renormalisation constant computed using a mass-independent renormalisation scheme \cite{Weinberg:1951ss}; $n$ is a light-like four-vector, $n^2=0$; $P$ is the kaon's four-momentum, $P^2=-m_K^2$ and $n\cdot P = -m_K$, with $m_K$ being the kaon's mass; and $(q_{\eta\bar\eta} = [q_\eta+q_{\bar\eta}]/2)$
\begin{equation}
\chi_K^P(q_\eta,q_{\bar\eta}) = S_s(q_\eta) \Gamma_K(q_{\eta\bar\eta};P) S_u(q_{\bar \eta})\,,
\label{chipi}
\end{equation}
is the kaon's Poincar\'e-covariant Bethe-Salpeter wave-function, with $\Gamma_K$ the Bethe-Salpeter amplitude, $S_{s,u}$ the dressed $s$- and $u$-quark propagators, which take the form
\begin{subequations}
\label{SgeneralN}
\begin{eqnarray}
S_{f=s,u}(q) &=& -i \gamma\cdot p\,\sigma_V^f(q^2) + \sigma_S^f(q^2)\\
&=& Z_f(q^2)/[i \gamma\cdot p + M_f(p^2)]\,,
\end{eqnarray}
\end{subequations}
and $q_\eta = q + \eta P$, $q_{\bar\eta} = q - (1-\eta) P$, $\eta\in [0,1]$.  Owing to Poincar\'e covariance, no observable can legitimately depend on $\eta$; i.e., the definition of the relative momentum.

With $\chi_K^P$ in hand, it is straightforward to generalise the procedure explained and employed in Ref.\,\cite{Chang:2013pqS}, and thereby obtain $\varphi_K(u)$ from Eq.\,\eqref{kaonPDA}.  One first computes the moments
\begin{equation}
\langle u_\Delta^m\rangle = \int_0^1 du \, (2u-1)^m \varphi_K(u)\,,
\end{equation}
which, using Eq.\,\eqref{kaonPDA}, can be obtained via
\begin{equation}
f_K (n\cdot P)^{m+1} \langle u_\Delta^m\rangle =
N_c {\rm tr}\,
Z_2 \! \int_{dq}^\Lambda \!\!
(2 n\cdot q_\eta-n\cdot P)^m \,\gamma_5\gamma\cdot n\, \chi_\pi^P(q_\eta,q_{\bar\eta})\,.
\label{phimom}
\end{equation}
Notably, beginning with an accurate form of $\chi_K^P$, arbitrarily many moments can be computed so that $\varphi_K(u)$ can reliably be reconstructed using the method we now describe.

Since the kaon is composed from valence-quarks with unequal current-masses, then $\varphi_K(u)\neq \varphi_K(1-u)$ and all moments produced by Eq.\,\eqref{phimom} are nonzero.  (The asymmetry disappears with the difference between current-quark masses: with mass degeneracy, the odd-$m$ moments vanish, as occurs, e.g., for the $\pi$-, $\rho$- and $\phi$-mesons \cite{Chang:2013pqS,Gao:2014bca}.)  It follows that one may write
\begin{subequations}
\begin{eqnarray}
\varphi_K(u) &= &  \varphi_K^E(u) + \varphi_K^O(u)\,,\\
\varphi_K^{E,O}(u) & = & (1/2)[\varphi_K(u)\pm \varphi_K(1-u)] \,. 
\end{eqnarray}
\end{subequations}
In this form, the nonzero moments of $\varphi_K^E(u)$ reproduce all the $m$-even moments of $\varphi_K$ and the nonzero moments of $\varphi_K^O(u)$ are the $m$-odd moments of $\varphi_K$.

Consider now that Gegenbauer polynomials of order $\alpha$, $\{C_n^{\alpha}(2 u -1)\,|\, n=0,\ldots,\infty\}$, are a complete orthonormal set on $u\in[0,1]$ with respect to the measure $[u (1-u)]^{\alpha_-}$, $\alpha_-=\alpha-1/2$.  They therefore enable reconstruction of any function defined on $u\in[0,1]$ that vanishes at the endpoints; and hence, with complete generality and to a level of accuracy defined by the summation upper bounds,
\begin{equation}
\label{PDAGalpha}
\varphi_K^{E,O}(u) \approx \,_m\varphi_K^{E,O}(u) \,,
\end{equation}
where
\begin{subequations}
\label{phimEO}
\begin{eqnarray}
\,_m\varphi_K^E(u) &= &
 N_{\bar \alpha} \, [u (1-u)]^{\bar\alpha_-}\!\!\!\!\!
\sum_{j=0,2,4,\ldots}^{\bar j_{\rm max}} a_j^{\bar\alpha} C_j^{\bar\alpha}(2 u -1)
\,, \quad \\
\,_m\varphi_K^O(u) &=& N_{\hat \alpha} \, [u (1-u)]^{\hat \alpha_-}\,
\sum_{j=1,3,\ldots}^{\hat{j}_{\rm max}+1} a_j^{\hat\alpha} C_j^{\hat\alpha}(2 u -1)\,,
\end{eqnarray}
\end{subequations}
$N_\alpha = \Gamma(2\alpha+1)/[\Gamma(\alpha+1/2)]^2$ and $a_0^{\bar\alpha} = 1$.  In general, $\bar\alpha \neq \hat\alpha$ because $\varphi_K^E(u)$ and $\varphi_K^O(u)$ are orthogonal components of $\varphi_K(u)$.

At this point, from a given set of $2 m_{\rm max}$ moments computed via Eq.\,\eqref{phimom}, the even and odd component-PDAs are determined independently by separately minimising
\begin{subequations}
\begin{eqnarray}
\varepsilon_m^{E} &=& \sum_{l=2,4,\ldots, 2 m_{\rm max}} |\langle  u_\Delta^l\rangle_{m}^E/\langle u_\Delta^l\rangle-1|\,,\\
\varepsilon_m^O &=& \sum_{l=1,3,\ldots, 2 m_{\rm max}-1} |\langle  u_\Delta^l\rangle_{m}^O/\langle u_\Delta^l\rangle-1|\,,
\end{eqnarray}
\end{subequations}
over the sets $\{ \bar\alpha, a_2, a_4, \ldots, a_{j_{\rm max}}\}$, $\{ \hat \alpha, a_1, a_3, \ldots, a_{j_{\rm max}+1}\}$,  where
\begin{equation}
\langle u_\Delta^l\rangle_{m}^{E,O} = \int_0^1 du\, (2u-1)^l \,_m\varphi_K^{E,O}(u)\,.
\label{endprocedure}
\end{equation}

This procedure acknowledges that at all empirically accessible scales the pointwise profile of PDAs is determined by nonperturbative dynamics \cite{Chang:2013pqS,Cloet:2013tta,Chang:2013epa,%
Cloet:2013jya,Chang:2013nia,Segovia:2013ecaS}; and hence they should be reconstructed from moments by using Gegenbauer polynomials of order $\alpha$, with the order $\alpha$ determined by the moments themselves, not fixed beforehand.  In the case of $\pi$-, $\rho$- and $\phi$-mesons, this procedure converges rapidly: $j_{\rm max}=2$ is sufficient \cite{Chang:2013pqS,Gao:2014bca}.

\medskip

\noindent\textbf{3.$\;$Results for the kaon twist-two PDA}.
We solved the $s$- and $u$- quark gap equations and the kaon Bethe-Salpeter equation numerically using the interaction in Ref.\,\cite{Qin:2011dd}.  The infrared composition of this interaction is deliberately consistent with that determined in modern studies of QCD's gauge sector \cite{Bowman:2004jm,Cucchieri:2011ig,Boucaud:2011ugS,Ayala:2012pb,%
Aguilar:2012rz,Strauss:2012dg}; and, in the ultraviolet, it preserves the one-loop renormalisation group behaviour of QCD so that, e.g., the dressed-quark mass-functions $M_{s,u}(p^2)= \sigma_S^{s,u}(p^2,\zeta^2)/\sigma_V^{s,u}(p^2,\zeta^2)$, are independent of the renormalisation point, which we choose to be $\zeta=2\,$GeV$\,=:\zeta_2$.  In completing the gap and Bethe-Salpeter kernels we employ two different procedures and compare their results: rainbow-ladder truncation (RL), detailed in App.\,A.1 of Ref.\,\cite{Chang:2012cc}, which is the most widely used DSE computational scheme in hadron physics, whose strengths and weakness are canvassed elsewhere \cite{Maris:2003vk,Chang:2011vu,Bashir:2012fs,Cloet:2013jya}; and the modern DCSB-improved kernels (DB) detailed in App.\,A.2 of Ref.\,\cite{Chang:2012cc}, which are the most refined kernels currently available \cite{Chang:2009zb,Chang:2010hb,Chang:2011ei,Cloet:2013jya}.  Both schemes are symmetry-preserving but the latter introduces essentially nonperturbative DCSB effects into the kernels, which are omitted in RL truncation and any stepwise improvement thereof.  The DB kernel is thus the more realistic.

\begin{figure}[t]

\centerline{\includegraphics[width=0.78\linewidth]{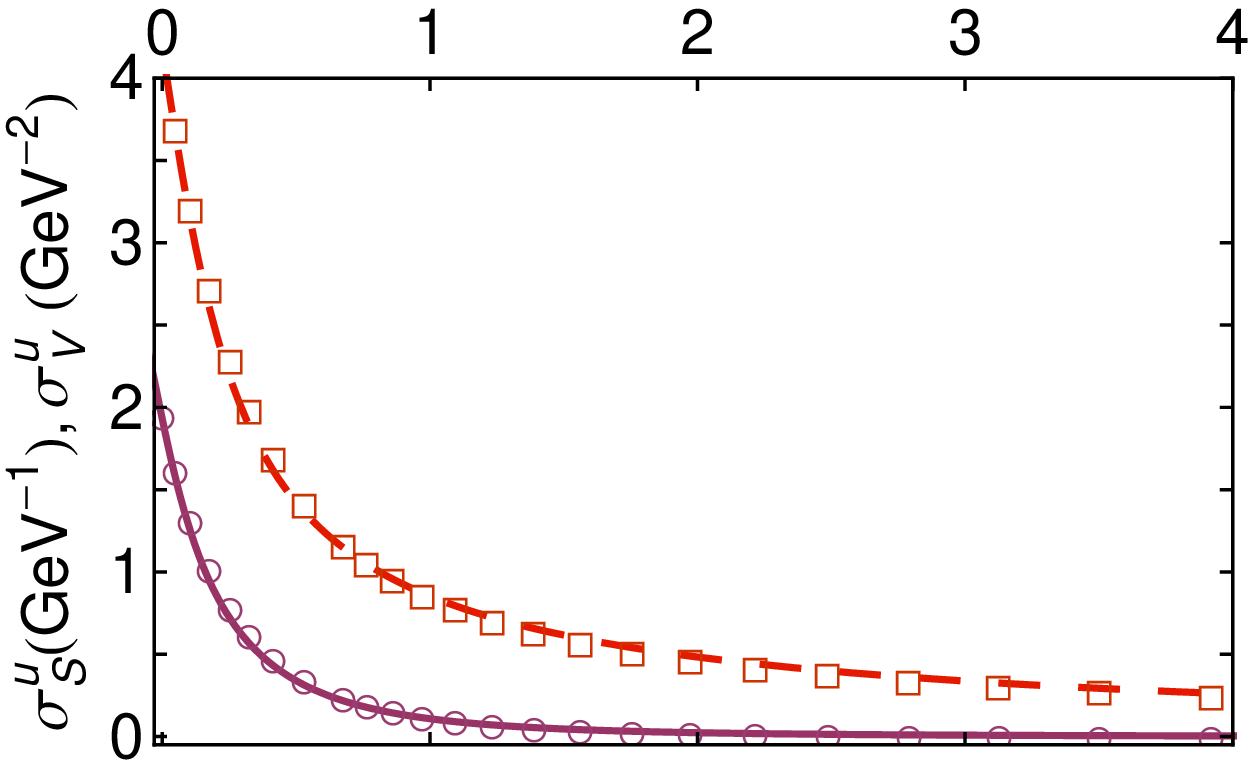}}

\vspace*{-8.3ex}
\centerline{\includegraphics[width=0.78\linewidth]{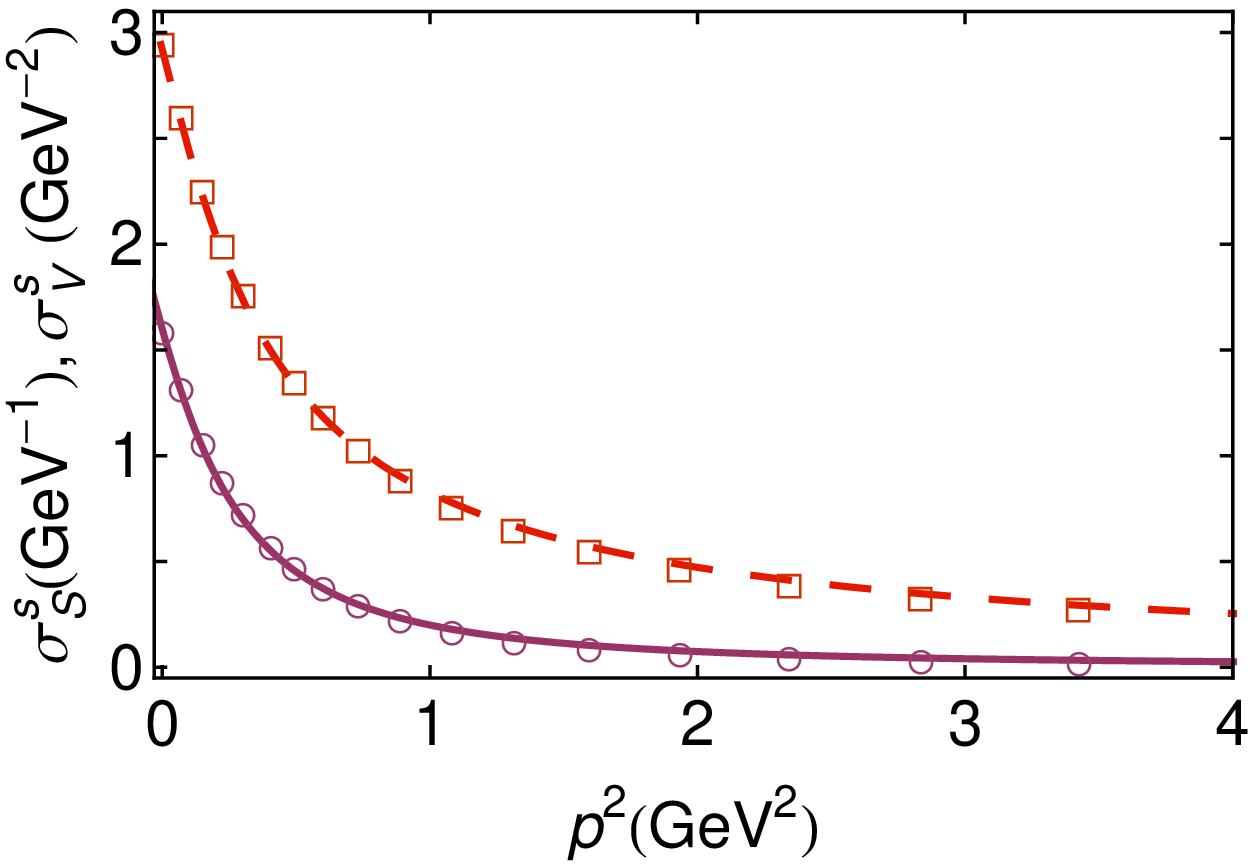}}

\caption{Functions characterising the dressed quark propagator in the DB truncation.
\emph{Upper panel}. $u/d$-quark functions, $\sigma_{S,V}^{u/d}(p^2)$ -- solution (open circles and squares, respectively) and interpolation functions (solid and long-dashed curves, respectively).
\emph{Lower panel}. $s$-quark\ functions, $\sigma_{S,V}^s(p^2)$.  Same legend.
\label{fig:Splot}}
\end{figure}

The gap and Bethe-Salpeter equation solutions are obtained as matrix tables of numbers.  Computation of the moments in Eq.\,\eqref{phimom} is cumbersome with such input, so we employ algebraic parametrisations of each array to serve as interpolations in evaluating the moments.  For the quark propagators, we represent $\sigma_{V,S}$ as meromorphic functions with no poles on the real $p^2$-axis \cite{Bhagwat:2002tx}, a feature consistent with confinement as defined through the violation of reflection positivity \cite{Gribov:1999ui,Krein:1990sf,%
Dokshitzer:2004ie,Roberts:2007ji,Chang:2011vu,Bashir:2012fs,%
Cloet:2013jya}.  Each scalar function in the kaon's Bethe-Salpeter amplitude is expressed via a Nakanishi-like representation \cite{Nakanishi:1963zz,Nakanishi:1969ph,Nakanishi:1971}, with parameters fitted to that function's first four $q\cdot P$ Chebyshev moments.  The quality of the description is illustrated via the dressed-quark propagator in Fig.\,\ref{fig:Splot}; and details are presented in the Appendix.

\begin{figure}[t]

\centerline{\includegraphics[width=0.8\linewidth]{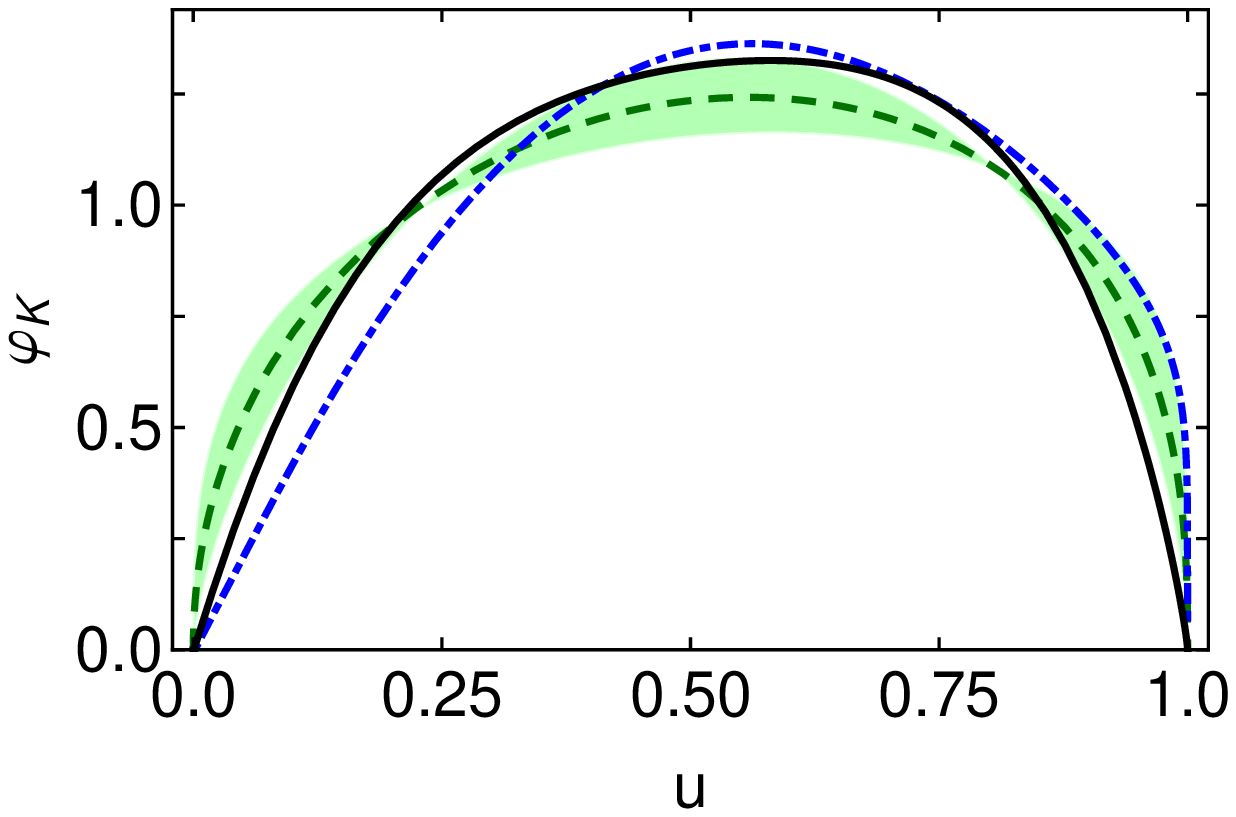}}

\centerline{\includegraphics[width=0.8\linewidth]{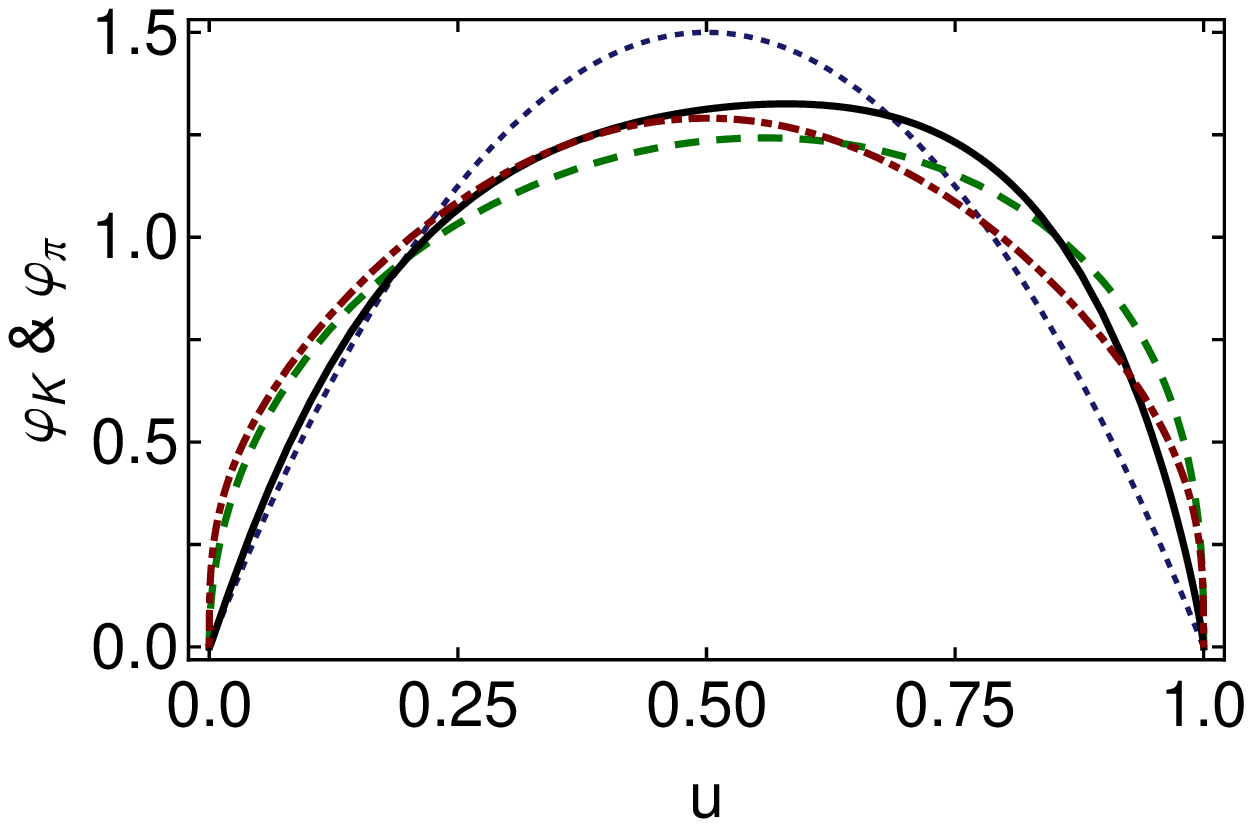}}

\caption{\emph{Upper panel} -- Kaon's twist-two valence-quark parton distribution amplitude.
Solid curve (black) -- result obtained with DB kernel;
dot-dashed curve (blue) -- RL kernel;
dashed line and band (green) -- result in Eq.\,\eqref{latticePDA}, inferred from two nontrivial moments obtained using lattice-QCD \cite{Segovia:2013ecaS}.
\emph{Lower panel} -- Comparison between DB kernel results for the PDAs of the kaon (solid, black) and pion (dot-dashed, red).  The ``best fit'' lattice-QCD result in Eq.\,\eqref{latticePDA} is also shown (dashed, green) along with the asymptotic PDA: $\varphi^{\rm asy}(u)=6 u (1-u)$ (dotted, dark-blue).
\label{fig:comp}}
\end{figure}

Using the interpolating spectral representations, it is straightforward to compute arbitrarily many moments of the kaon's PDAs via Eqs.\,\eqref{phimom}. We typically employ $2 m_{\rm max}=50$.  The pointwise forms of the PDAs are then reconstructed via the ``Gegenbauer-$\alpha$'' procedure described in connection with Eqs.\,\eqref{PDAGalpha}--\eqref{endprocedure} above.  Again, the procedure converges rapidly, so that results obtained with $j_{\rm max}=2$ produce $\epsilon_m^{E,O} < 1$\%.

Our results, computed at the renormalisation scale $\zeta_2$ and depicted in Fig.\,\ref{fig:comp}, are described by:
\begin{eqnarray}
\label{phiKA}
\varphi_K(u) = \,_m\varphi_K^E(u)+ \,_m\varphi_K^O(u)
\end{eqnarray}
with the functions defined in Eqs.\,\eqref{phimEO} and
\begin{equation}
\label{phiKB}
\begin{array}{lccccc}
   & \bar\alpha & \hat \alpha & a_2^{\bar\alpha} & a_1^{\hat\alpha} & a_3^{\hat\alpha} \\
{\rm RL} & 0.68 & 0.65 & -0.32\phantom{0} & 0.27& 0.054 \\
{\rm DB} & 1.42 & 1.14 & \phantom{-}0.074 & 0.076 & 0.011\\
\end{array}\,.
\end{equation}

To assist in making comparisons with results obtained using other methods, we list the lowest six moments computed using Eqs.\,\eqref{phiKA}, \eqref{phiKB} in Table~\ref{momentsRLDB}.  In considering Table~\ref{momentsRLDB}, it should be borne in mind that only our study and those using lattice-QCD can unambiguously determine the scale at which the calculation is valid: the lattice results were also obtained at $\zeta_2$. Sum-rules studies, on the other hand, are thought to be defined at some vaguely determined ``typical hadronic scale'', which cannot realistically be known to within better than a factor of two.

\begin{table}[t]
\caption{Moments ($u_\Delta=2u-1$) of the $K$-meson PDA computed using Eqs.\,\eqref{phiKA} and \eqref{phiKB}, compared with selected results obtained elsewhere: %
Refs.\,\cite{Braun:2006dg,Arthur:2010xf}, lattice-QCD;
Ref.\,\cite{Segovia:2013ecaS}, analysis of lattice-QCD results in Ref.\,\cite{Arthur:2010xf};
Refs.\,\cite{Khodjamirian:2004ga,Braun:2004vf,Ball:2005vx,Ball:2006fz}, compilation of results from QCD sum rules;
and Ref.\,\cite{Brodsky:2008pg}, holographic soft-wall \emph{Ansatz} for the kaon's light-front wave function.
%
%
We also list values obtained with $\varphi = \varphi^{\rm asy}$, Eq.\,\eqref{momUV}, and $\varphi = \varphi_{\rm ms}$, Eq.\,\eqref{momms}, because they represent lower and upper bounds, respectively, for concave distribution amplitudes.
\label{momentsRLDB}
}
\begin{tabular*}
{\hsize}
{
l|
l@{\extracolsep{0ptplus1fil}}
l@{\extracolsep{0ptplus1fil}}
l@{\extracolsep{0ptplus1fil}}
l@{\extracolsep{0ptplus1fil}}
l@{\extracolsep{0ptplus1fil}}
l@{\extracolsep{0ptplus1fil}}}\hline
 $\langle u_\Delta^m \rangle$    & $m=1$ & $2$ & $3$ & $4$ & $5$ & $6$\\\hline

RL & 0.11\phantom{0} & 0.24 & 0.064 & 0.12 & 0.045 & 0.076 \\
DB & 0.040 & 0.23 & 0.021 & 0.11 & 0.013 & 0.063\\\hline
%
\cite{Braun:2006dg} & 0.027(2) & 0.26(2) &   &   &   & \\
\cite{Arthur:2010xf} 
    &  0.036(2)  & 0.26(2) &   &   &   & \\
\cite{Segovia:2013ecaS}
    &  0.036(2)  & 0.26(2) & 0.020(2)  & 0.13(2)  & 0.014(2)  & 0.085(15)\\ \hline
\cite{Khodjamirian:2004ga,Braun:2004vf,Ball:2005vx,Ball:2006fz}
    & 0.04(8) & &   &   &   & \\
\cite{Brodsky:2008pg}
    & 0.04(2) & 0.24(1) &   &   &   & \\\hline
%
$\varphi=\varphi_{\rm ms}$  & 0.33 & 0.33 & 0.2 & 0.2 & 0.14 & 0.14 \\
$\varphi=\varphi^{\rm asy}$ & 0 & 0.2 & 0 & 0.086 & 0 & 0.048 \\\hline
\end{tabular*}
\end{table}

In Table~\ref{momentsRLDB} we also list all moments of the kaon's PDA that can be computed with contemporary algorithms via numerical simulations of lattice-regularised QCD \cite{Braun:2006dg,Arthur:2010xf}.  Working with the most recent results \cite{Arthur:2010xf} and using the method introduced in Refs.\,\cite{Chang:2013pqS,Cloet:2013tta,Chang:2013epa}, which is founded in Bayesian analysis, one can obtain a reliable pointwise approximation to the kaon's PDA from this limited information.  The result is a concave function, represented by \cite{Segovia:2013ecaS}
\begin{equation}
\label{latticePDA}
\varphi_K(u) = N_{\alpha\beta}\, u^\alpha (1-u)^\beta\,,
\;  \alpha_{su} = 0.48^{+0.19}_{-0.16}\,,\;
\beta_{su} = 0.38^{+0.17}_{-0.15},
\end{equation}
where $N_{\alpha\beta}=1/B(1+\alpha,1+\beta)$.  This function and the associated error band are depicted in Fig.\,\ref{fig:comp}.

It is useful to provide limits on the allowed values of the moments in Table~\ref{momentsRLDB}.  In the present context, two extremes are possible.  As the scale $\zeta\to\infty$, $\varphi_K(u) \to \varphi^{\rm asy}(u)$, so the moments of $\varphi^{\rm asy}(u)$ provide a lower bound for any reasonable PDA:
\begin{equation}
\label{momUV}
\int_0^1du\, (2u-1)^m\,\varphi^{\rm asy}(u) = \frac{3 \left(1+(-1)^m\right)}{2 (m+1) (m+3)}\,.
\end{equation}
On the other hand, the most skewed concave distribution amplitude possible is obtained via
\begin{equation}
\label{maxskew}
\varphi_{\rm ms}(u):= \lim_{\alpha\to 1, \beta\to 0}N_{\alpha\beta} u^\alpha (1-u)^\beta = 2\,u\,;
\end{equation}
and hence the moments of $\varphi_{\rm ms}(u)$ provide an upper bound:
\begin{equation}
\label{momms}
\int_0^1du\, (2u-1)^m\,\varphi_{\rm ms}(u) = \frac{2 m+3+(-1)^m}{2 (m+1) (m+2)}\,.
\end{equation}
Notably, the even moments obtained with Eq.\,\eqref{momms} are those of the distribution amplitude $\varphi(u) = \,$constant, the odd moments of which vanish.  We list the limiting moments in Table~\ref{momentsRLDB}.\footnote{The association of Eq.\,\eqref{maxskew} with a maximally skewed distribution is further clarified by noting that this PDA is produced by using $\rho_0(\alpha)\to \delta(1-\alpha)=:\rho_{\rm ms}(\alpha)$ in Eq.\,\eqref{fit} and setting $n_0=1$, $U_1=0=U_2$.  With this choice of spectral function, all the bound-state's momentum is plainly lodged with the valence quark.
}

There are a number of important messages to be read from Fig.\,\ref{fig:comp}. The upper panel shows that the kaon distribution is skewed: the RL amplitude peaks at $u=0.56$; DB at $u=0.58$; and the result inferred from lattice-QCD peaks at $u=0.56^{+0.02}_{-0.01}$.  In a meson constituted from valence-quarks with equal current-mass, the distribution amplitude is symmetric and peaks at $u=1/2$.  The unambiguous conclusion is that, on the light-front, the $s$-quark carries more of the kaon's momentum than the $\bar u$ quark.

This 12-16\% shift in peak location is a quantitative measure of SU$(3)$-flavour-symmetry breaking in hadrons.  It is comparable with
the 15\% shift in the peak of the kaon's valence $s$-quark parton distribution function, $s_v^K(x)$, relative to $u_v^K(x)$ \cite{Nguyen:2011jy} and the ratio of neutral- and charged-kaon electromagnetic form factors measured in $e^+e^-$ annihilation at $s_U=17.4\,$GeV$^2$ \cite{Seth:2013eaa}: $|F_{K_S K_L}(s_U)|/|F_{K_- K_+}(s_U)|\approx 0.12$.
By way of context, it is notable that the ratio of $s$-to-$u$ current-quark masses is approximately $27$ \cite{Beringer:1900zz}, whereas the ratio of nonperturbatively generated Euclidean constituent-quark masses is typically $1.5$ \cite{Chen:2012qr} and the ratio of leptonic decay constants $f_K/f_\pi \approx 1.2$ \cite{Beringer:1900zz}.  Both latter quantities are equivalent order parameters for DCSB.
Moreover, a DSE-based computation of leptonic decay constant ratios yields $f_{B_s}/f_B = 1.2$ \cite{Ivanov:2007cw}, in accord with a recent result from unquenched lattice-QCD $f_{B_s}/f_B=1.22(8)$ \cite{Christ:2014uea}, and the same DSE framework produces $f^+_{BK}(0)/f^+_{B\pi}(0)=1.21$ for the ratio of $B\to K,\pi$ semileptonic transition form factors at the maximum recoil point, a value that is typical for estimates of this quantity: the results in Refs.\,\cite{Melikhov:1997wp,Melikhov:2001zv,Faessler:2002ut,Ball:2004ye,%
Khodjamirian:2006st,Ebert:2006nz,Lu:2007sg} may be summarised as $f^+_{BK}(0)/f^+_{B\pi}(0)=1.26(5)$.
It is therefore apparent that the flavour-dependence of DCSB rather than explicit chiral symmetry breaking is measured by the skewness of $\varphi_K(u)$: SU$(3)$-flavour-symmetry breaking is far smaller than one might na\"ively have expected because DCSB impacts heavily on $u,d$- and $s$-quarks.

Focusing on the DSE results in the upper panel of Fig.\,\ref{fig:comp}, one observes that the RL PDA is more skewed than the DB result; viz., the RL truncation allocates a significantly larger fraction of the kaon's momentum to its valence $s$-quark.  This feature is also highlighted by comparing the RL and DB results for the moments in Table~\ref{momentsRLDB}: the $m=1,3,5$ RL moments are noticeably larger than the odd moments obtained with the DB kernel; and all RL moments are closer to the upper bound expressed in Eq.\,\eqref{momms}.
This is readily understood.
RL-kernels ignore DCSB in the quark-gluon vertex.  Therefore, to describe a given body of phenomena, they must shift all DCSB strength into the infrared behaviour of the dressed-quark propagator, whilst nevertheless maintaining perturbative behaviour for $p^2>\zeta_2^2$.  This requires $M_{s,u}(p^2)$ to be unnaturally large at $p^2=0$ and then drop quickly with increasing $p^2$, behaviour which influences $\varphi_K(u)$ via the Bethe-Salpeter equation.
In contrast, the DB-kernel builds DCSB into the quark-gluon vertex and its impact is therefore shared between more elements of a calculation.  Hence  smaller values of $M_{s,u}(p^2=0)$ are capable of describing the same body of phenomena; and these dressed-masses need fall less rapidly in order to reach the asymptotic limits they share with the RL self-energies.  The DB kernel therefore produces a more balanced expression of DCSB's impact on a meson's Bethe-Salpeter wave function and hence the PDA derived therefrom provides a more realistic expression of DCSB-induced skewness: it provides the most realistic result.
The behaviour of the even moments has a similar origin.

The preceding observations enable us to highlight a final feature of the upper panel in Fig.\,\ref{fig:comp}; namely, the agreement between the DB result for the kaon's PDA and that inferred from lattice-QCD.  The DB result is determined by one parameter, whose role is to express the infrared strength of the gap equation's kernel and whose value was chosen to reproduce the value of $f_\pi$, the pion's leptonic decay constant.
The same DB kernel describes a wide range of hadron physics observables \cite{Chang:2013pqS,Chang:2011ei} and no parameters were varied in order to produce the results described herein.  Therefore, the match between the DSE-DB result and that inferred from lattice-QCD suggests strongly that we have now arrived at a reliable form of the kaon's PDA and an understanding of flavour symmetry breaking therein.  

The lower panel of Fig.\,\ref{fig:comp} facilitates a comparison between the kaon's twist-two PDA and that obtained for the pion using the same kernel \cite{Chang:2013pqS}.  Plainly, at the scale $\zeta_2$ the kaon's PDA possesses dilation of the same magnitude as that present in $\varphi_\pi(u)$: both are significantly broader than the asymptotic PDA for mesons, $\varphi^{\rm asy}(u)$.  This hardness of the distributions at an hadronic scale is a direct expression of DCSB.  As shown elsewhere \cite{Cloet:2013tta,Cloet:2013jya,Segovia:2013ecaS}, it persists to energy scales $\zeta$ that exceed those available even at the large hadron collider.  Consequently, $\varphi^{\rm asy}(u)$ cannot be used to obtain reliable estimates for observable quantities at any energy scale that is currently conceivable in connection with terrestrial facilities.  Instead, the DCSB-dilated amplitudes should be used to obtain such information.

As an illustration, consider the ratio of kaon and pion electromagnetic form factors, which has been measured in $e^+ e^-$ annihilation on a large domain, with an upper bound of $s_U=17.4\,$GeV$^2$ \cite{Seth:2012nnS}: $|F_K(s_U)|/|F_\pi(s_U)| = 0.92(5)$.  At leading-order and leading twist, perturbative QCD (pQCD) predicts \cite{Lepage:1979zb,Farrar:1979aw,Efremov:1979qk,Lepage:1980fj}:
\begin{align}
\label{pionUV}
\exists Q_0 > & \Lambda_{\rm QCD} \; |\;   Q^2 F_P(Q^2) \stackrel{Q^2 > Q_0^2}{\approx} 16 \pi \alpha_s(Q^2)  f_P^2 \mathpzc{w}_{\varphi_P}^2,
\end{align}
with $Q^2$ spacelike and
\begin{subequations}
{\allowdisplaybreaks
\begin{align}
\label{wphi}
\mathpzc{w}_{\varphi_P}^2 &=
    e_{q_1} \mathpzc{w}_{\varphi_{q_1}}^2
+   e_{\bar q_2} \mathpzc{w}_{\varphi_{q_2}}^2\,,\\
\mathpzc{w}_{\varphi_{q_1}} &= \frac{1}{3} \int_0^1 du\, \frac{1}{1-u}\, \varphi_P(x)\,,\;
\mathpzc{w}_{\varphi_{q_2}}=\frac{1}{3} \int_0^1 du\, \frac{1}{u}\, \varphi_P(u)\,,
\end{align}}
\end{subequations}
\hspace*{-0.5\parindent}where $\alpha_s(Q^2)$ is the strong running coupling, $f_P$ is the meson's leptonic decay constant and $\varphi_P(x)$ is its PDA, and $e_{q_1,\bar q_2}$ are, respectively, the electric charges of the valence-quark and -antiquark in the meson: $e_{q_1}^K=e_s$, $e_{q_1}^\pi=e_d$, $e_{q_2}^{K,\pi}=e_{\bar u}$.  Using our DB-kernel results for $\varphi_{K,\pi}(u)$ and the one-loop expression for $\alpha_s(Q^2)$, with $\Lambda_{\rm QCD}=0.234\,$GeV and $N_f=4$ \cite{Qin:2011dd}, we employ the one-loop evolution equations \cite{Efremov:1979qk,Lepage:1980fj} to express our amplitudes at $\zeta_E^2=17.4\,$GeV$^2$, and therewith obtain
\begin{equation}
F_K(\zeta_E^2)/F_\pi(\zeta_E^2) = 1.23\,.
\end{equation}
This prediction follows from the computed values: $\omega_{q_1=s}^K=1.21$, $\omega_{q_2=\bar u}^K=1.0$, $\omega^\pi=1.17$, which expose a 17\% SU$(3)$-flavour-symmetry breaking effect at $\zeta_E$; and it agrees with the value inferred from experiment to better than 30\%, despite the experiment being performed at timelike momenta.  The claim \cite{Seth:2013eaa} of a $9\sigma$ (factor of two) disagreement in this ratio between pQCD and experiment is thus revealed to be a misapprehension, arising because the expected result was based on a mistaken assumption that $\varphi^{\rm asy}$ should provide estimates relevant to contemporary experiment.  This repeats a pattern predicted for the pion form factor itself \cite{Chang:2013nia}, in which parton-model scaling and scaling violations are apparent on $Q^2\gtrsim 8\,$GeV$^2$ but the normalisation is set by nonperturbative DCSB dynamics.

We would like to remark that whilst agreement between experiment and theory for the ratio is satisfactory at $s_U$, a puzzle remains with the normalisation of $F_{K,\pi}(s_U)$ \cite{Holt:2012gg}.  This is highlighted by a comparison between the computed value of $F_\pi(\zeta_E^2)=0.42/\zeta_E^2$ \cite{Chang:2013nia} and $|F_\pi(s_U)|=0.84(5)/s_U$ reported in Ref.\,\cite{Seth:2012nnS}.  The computation in Ref.\,\cite{Chang:2013nia} agrees with all available, reliable spacelike data, and the calculated value of $F_\pi(\zeta_E^2)$ is a factor of four larger than the result obtained from Eq.\,\eqref{pionUV} using $\varphi^{\rm asy}$.  It is nevertheless still a factor of two smaller than the stated timelike experimental value.

\medskip

\noindent\textbf{4.$\;$Conclusion}.
We described the first Dyson-Schwinger equation (DSE) computation of the valence-quark (twist-two parton) distribution amplitude for a bound-state constituted from quarks with unequal current masses; namely, the kaon.  In this case, the PDA is broad, concave and skewed; i.e., asymmetric, with the peak located at $u=0.56$-$0.58$.  These features are a clear and accurate expression of SU$(3)$-flavour-symmetry breaking in hadron physics.  They show that: the heavier quark in the kaon carries more of the bound-state's momentum; and the scale of flavour-symmetry breaking is nonperturbative in origin.  Indeed, the same can be said for the PDA's $u$-dependence at any accessible energy scale.  Our results are consistent with those inferred from numerical simulations of lattice-regularised QCD; and this confluence suggests strongly that the kaon (and pion) PDA described above should serve as the basis for future attempts to access CP violation in the Standard Model.

It is worth reiterating that there are a number of advantages in using the DSE approach in studies such as this.  For example, the framework preserves the one-loop renormalisation group behaviour of QCD, so that current-quark masses have a direct connection with the parameters in QCD's action and the dressed-quark mass-functions, $M_{s,u}(p^2)$, are independent of the renormalisation point.  Unlike other approaches to nonperturbative phenomena in continuum QCD, the renormalisation point can be fixed unambiguously, as in lattice-QCD: it is not a parameter to be identified with some poorly determined ``typical hadronic scale.''  Moreover, one is not just restricted to estimating a few low-order moments of the PDA.  In working in the continuum and computing Bethe-Salpeter wave functions directly, the DSEs enable one to deliver a prediction for the pointwise behaviour of the PDA on the full domain $u\in [0,1]$.  Importantly, that prediction is parameter-free and unifies the kaon's PDA with a diverse range of apparently distinct phenomena.

A coherent picture is now emerging.  Modern DSE studies predict PDAs for light-quark mesons that are broad concave functions.  The dilation with respect to the asymptotic PDA is a clean expression of dynamical chiral symmetry breaking (DCSB) on the light front.  Notably, where a comparison is possible, the DSE results are consistent with those determined via contemporary numerical simulations of lattice-regularised QCD.  A new paradigm thus presents itself, from which it follows that at energy scales accessible with existing and foreseeable facilities, one may arrive at reliable expectations for the outcome of experiments by using these broad, concave PDAs in the leading-order, leading-twist formulae for hard exclusive processes.  Following this procedure, any discrepancies will be significantly smaller than those produced by using the asymptotic PDA in such formulae and the magnitude of the disagreement will provide a good estimate of the size of higher-order, higher-twist effects.

\medskip

\noindent\textbf{Acknowledgments}.
We benefited from insightful comments by I.\,C.~Clo\"et, B.~El-Bennich, R.\,J.~Holt, G.~Krein, J.~Segovia and A.\,W.~Thomas; and from the opportunity to participate (LC, CDR, PCT) in the workshops ``Many Manifestations of Nonperturbative QCD under the Southern Cross'', Ubatuba, S\~ao Paulo, and (CDR, PCT) the ``2$^{\rm nd}$ Workshop on Perspectives in Nonperturbative QCD'' at IFT-UNESP, S\~ao Paulo, during both of which a substantial body of this work was completed.
CDR acknowledges support from an \emph{International Fellow Award} from the Helmholtz Association; and
research otherwise supported by:
the National Natural Science Foundation of China (grant nos.\ 11275097
and 11274166); the National Basic Research Program of China (grant no.\ 2012CB921504); the Research Fund for the Doctoral Program of Higher Education (China, grant no. 2012009111002);
University of Adelaide and Australian Research Council through grant no.~FL0992247;
Department of Energy, Office of Nuclear Physics, contract no.~DE-AC02-06CH11357;
For\-schungs\-zentrum J\"ulich GmbH;
and
National Science Foundation, grant no.\ NSF-PHY1206187.

\medskip

\appendix

\setcounter{equation}{0}
\setcounter{table}{0}
\renewcommand{\theequation}{A\arabic{equation}}
\renewcommand{\thetable}{A.\arabic{table}}

\noindent\textbf{Appendix}.
Here we describe the interpolations used in our evaluation of the moments in Eq.\,\eqref{phimom}.  There are two sets of results to consider; viz., those obtained in RL truncation and those produced by DB truncation.  The interaction in Ref.\,\cite{Qin:2011dd} has one parameter $m_g^3 := D\omega$ because with $m_g = \,$constant, light-quark observables are independent of the value of $\omega \in [0.4,0.6]\,$GeV.  We use $\omega =0.5\,$GeV.

In RL truncation, with $m_g = 0.82\,$GeV and renormalisation point invariant current-quark masses $\hat m_{u/d}=6.2\,$MeV, $\hat m_s = 160\,$MeV, which correspond to the following one-loop evolved masses $m_{u/d}^{\zeta=2\,{\rm GeV}} = 4.3\,$MeV, $m_{s}^{\zeta=2\,{\rm GeV}} =110\,$MeV, we obtain $m_\pi=0.14\,$GeV, $f_\pi=0.093\,$GeV and $m_K=0.49$GeV, $f_K=0.11\,$GeV.

Using the DB truncation with $m_g = 0.55\,$GeV, we obtain $m_\pi=0.14\,$GeV, $m_K=0.50$GeV from renormalisation point invariant current-quark masses $\hat m_{u/d}=4.4\,$MeV, $\hat m_s = 90\,$MeV, which yield $m_{u/d}^{\zeta=2\,{\rm GeV}} = 3.0\,$MeV, $m_{s}^{\zeta=2\,{\rm GeV}} =62\,$MeV and produce the following values of the dressed-quark mass $M_u(\zeta_2)=4.3\,$MeV, $M_s(\zeta_2)=89\,$MeV, which are in fair agreement with modern lattice estimates \cite{Carrasco:2014cwa}.

In interpolating the results from either truncation, the dressed-quark propagators are represented as \cite{Bhagwat:2002tx}
\begin{equation}
S_f(p) = \sum_{j=1}^{j_m}\bigg[ \frac{z_j^f}{i \gamma\cdot p + m_j^f}+\frac{z_j^{f\ast}}{i \gamma \cdot p + m_j^{f\ast}}\bigg], \label{Spfit}
\end{equation}
with $\Im m_j \neq 0$ $\forall j$, so that $\sigma_{V,S}$ are meromorphic functions with no poles on the real $p^2$-axis, a feature consistent with confinement \cite{Bashir:2012fs}.  We find that $j_m=2$ is adequate; and the interpolation parameters are listed in Table~\ref{paramsquark}.

\begin{table}[t]
\caption{Representation parameters. Eq.\,\protect\eqref{Spfit} -- the pair $(x,y)$ represents the complex number $x+ i y$.  (Dimensioned quantities in GeV).
\label{paramsquark}
}
\begin{center}
\begin{tabular*}
{\hsize}
{
l|@{\extracolsep{0ptplus1fil}}
c@{\extracolsep{0ptplus1fil}}
c@{\extracolsep{0ptplus1fil}}
c@{\extracolsep{0ptplus1fil}}
c@{\extracolsep{0ptplus1fil}}
c@{\extracolsep{0ptplus1fil}}}\hline
 RL & $z_1$ & $m_1$  & $z_s$ & $m_2$ \\

 $u$& $(0.38,0.71)$ & $(0.71,0.22)$ & $(0.14,0)$ & $(-0.78,0.75)$ \\
 $s$& $(0.45,0.15)$ & $(0.72,0.29)$ & $(0.16,0.01)$ & $(-1.45,0.74)$ \\
 DB & $z_1$ & $m_1$  & $z_s$ & $m_2$ \\
 $u$& $(0.42,0.24)$ & $(0.44,0.19)$ & $(0.13,0.07)$ & $(-0.76,0.60)$ \\
 $s$& $(0.43,0.30)$ & $(0.55,0.22)$ & $(0.12,0.11)$ & $(-0.83,0.42)$ \\\hline
\end{tabular*}
\end{center}
\end{table}

\begin{table}[t]
\caption{Representation parameters associated with Eqs.\,\eqref{BSK}--\eqref{rho}. (Dimensioned quantities in GeV.  Omitted quantities are zero or unused.)
\label{BSAparameters}
}
\begin{center}
\begin{tabular*}
{\hsize}
{
r|@{\extracolsep{0ptplus1fil}}
c@{\extracolsep{0ptplus1fil}}
c@{\extracolsep{0ptplus1fil}}
c@{\extracolsep{0ptplus1fil}}
c@{\extracolsep{0ptplus1fil}}
c@{\extracolsep{0ptplus1fil}}
c@{\extracolsep{0ptplus1fil}}}\hline
 RL & $E_0$ & $E_1$  & $F_0$ & $F_1$ & $G_0$ & $G_1$ \\\hline
 $\nu_{0}$& $-0.71$ & $0.17$ & $1.33$ & 5.62 & $1.0$ & -0.1 \\
 $\nu_1$  &         &        &        &      & $-0.7$&      \\
 $\nu_{2}$& $1.0$ & $0.0$ & $0.0$ & $0.0$ & $0.0$ & $0.0$\\
 $U_0$ & $1.0$ & $0.7$& $0.42$  & 0.21 & 0.0 & $0.28$ \\
 $U_1$ &       &      &         &      & 0.25&        \\
 $10^3 U_2$& 6.83 & 0.36& 0.90  & 0.01 & -0.01 & 0.70 \\
 $n_0$& $5$ & 8& $5$ & 8 & 10 & 6 \\
 $n_1$&     &  &     &   & 12 &   \\
 $n_2$& $1$ &$ 2$& $1$ &$ 2$ &$ 2$ &$ 2$\\
 $\Lambda$& $1.8$ & $2.0$& $1.5$ & 1.6 & 2.1 & $1.5$ \\\hline
 DB & $E_0$ & $E_1$  & $F_0$ & $F_1$ & $G_0$ & $G_1$ \\\hline
$\nu_{0}$& $-0.54$ & $-0.1$ & $-0.01$ & 1.6 & 1.5 & $3.0$\\
$\nu_{1}$& $-0.7$ & $-0.4$& $-0.7$ & 0.8 &   & 3.0 \\
$\nu_{2}$& $1.0$ & $0.0$ & $0.0$ & $0.0$ & $0.0$ & $0.0$\\
$U_0$ & $1.0$ & $0.22$& $0.56$ & 0.11 & -0.058 & $0.12$ \\
$U_1$ & -2.0 & -0.5& -0.3  & -0.65 &    & -1.5\\
$10^2U_2$& 2.5 & 0.052 & 0.39& 0.001 & 0.049 & -0.60 \\
$n_0$& $4$ & 8& $4$ & 10 & 5 & 8 \\
$n_1$& $5$ & $12$& $6$ & 12 &   & 10 \\
$n_2$& $1$ &$ 2$& $1$ &$ 2$ & 2 & 2\\
$\Lambda$& $1.35$ & $1.7$& $1.2$ & 1.45 & 0.8 & $1.1$ \\\hline
\end{tabular*}
\end{center}
\end{table}

The kaon's Bethe-Salpeter amplitude has the form $(\ell = q_{\eta\bar\eta})$
\begin{eqnarray}
\nonumber
\lefteqn{\Gamma_{K}(\ell;P) = \gamma_5
\big[ i E_{K}(\ell;P) + \gamma\cdot P F_{K}(\ell;P)   }\\
&&  \quad\quad  + \gamma\cdot \ell \, G_{K}(\ell;P) + \sigma_{\mu\nu} \ell_\mu P_\nu H_{K}(\ell;P) \big]. 
\label{BSK}
\end{eqnarray}
As the kaon's valence-quarks are not degenerate in mass, each scalar function in Eq.\,\eqref{BSK} has the following decomposition
\begin{eqnarray}
\mathcal{F}(\ell;P)=\mathcal{F}_0(\ell;P)+\ell \cdot P \,\mathcal{F}_1(\ell;P)\,,
\label{decom}
\end{eqnarray}
with $\mathcal{F}_{0,1}\neq 0$ and even under $(\ell\cdot P) \to (-\ell\cdot P)$.  The following forms are flexible enough to allow a satisfactory representation of the numerical solutions to the Bethe-Salpeter equations:
\begin{eqnarray}
\nonumber \mathcal{F}_j(\ell,P)&=&
\int_{-1}^{1}d\alpha \, \rho_0(\alpha) \frac{(U_0 -U_1-U_2)\Lambda_j^{2 n_0}}{(\ell^2+\alpha\, \ell \cdot P+\Lambda_j^2)^{n_0}} \nonumber \\
\nonumber
&& +\int_{-1}^{1}d\alpha \, \rho_1(\alpha) \frac{U_1\Lambda_j^{2 n_1}}{(\ell^2+\alpha\, \ell \cdot P+\Lambda_j^2)^{n_1}}
\\
&&+\int_{-1}^{1}d\alpha \, \rho_2(\alpha) \frac{U_2\Lambda_j^{2 n_2}}{(\ell^2+\alpha\, \ell \cdot P+\Lambda_j^2)^{n_2}}\,,
\label{fit}
\end{eqnarray}
where
\begin{equation}
\rho_i(\alpha) = \frac{\Gamma(\nu_i+\frac{3}{2})}{\sqrt{\pi}\Gamma(\nu_i+1)}(1-\alpha^2)^{\nu_i}.
\label{rho}
\end{equation}

Values for the interpolation parameters in Eqs.\,\eqref{fit}, \eqref{rho} are determined via a least-squares fit to the Chebyshev moments
\begin{equation}
{\cal F}^{n}_{1,2}(\ell^2) = \frac{2}{\pi}\int_{-1}^{1}\!dx\, \sqrt{1-x^2} \,{\cal F}_E^{1,2}(\ell;P) U_n(x)\,,
\end{equation}
with $n=0,2$, where $U_n(x)$ is an order-$n$ Chebyshev polynomial of the second kind, and $i x = \hat \ell\cdot \hat P$, with $\hat \ell^2=1$ and $\hat P^2=-1$.  The resulting parameter values are listed in Table~\ref{BSAparameters}.  N.B.\ We have not included the overall multiplicative factor resulting from canonical normalisation of $\Gamma_K$; and the function $H$ is omitted because it does not have a noticeable effect on our results.



\end{document}